\documentclass[reprint,aps,prb,floatfix,amsmath,amssymb,longbibliography,superscriptaddress]{revtex4-1}
\usepackage{amsmath, amssymb, bm, graphicx}
\usepackage{xcolor}
\usepackage[normalem]{ulem}
\usepackage{mathtools}
\usepackage[stable]{footmisc}
\usepackage[colorlinks, linkcolor= blue, citecolor = blue, urlcolor=blue]{hyperref}
\usepackage{multirow}
\usepackage{array}
\usepackage{chemformula}
\usepackage{physics}
\def\be{\begin{equation}}
\def\ee{\end{equation}}
\def \bea{\begin{eqnarray}}
\def \eea{\end{eqnarray}}
\def \nn{\nonumber}

\begin{document}
\title{Diagnosing the origin of quantum oscillation beating in graphene} 

\author{Akash Adhikary}
\affiliation{Department of Physics, Indian Institute of Technology Kanpur, Kanpur 208016, India}
\author{Sunit Das}
\affiliation{Department of Physics, Indian Institute of Technology Kanpur, Kanpur 208016, India}
\author{Divya Sahani} 
\affiliation{Department of Physics, Indian Institute of Science, Bangalore 560012, India}
\affiliation{Chair of Electronic Devices, RWTH Aachen University, Aachen, Germany}
\author{Aveek Bid}
\email{aveek@iisc.ac.in}
\affiliation{Department of Physics, Indian Institute of Science, Bangalore 560012, India}
\author{Amit Agarwal}
\email{amitag@iitk.ac.in}
\affiliation{Department of Physics, Indian Institute of Technology Kanpur, Kanpur 208016, India}

\begin{abstract}
Magnetic quantum oscillations are usually periodic in inverse magnetic field, and their amplitude can show beating when two nearby frequencies interfere. In graphene-based hexagonal systems, such beating can arise from strain-induced pseudomagnetic fields, unequal valley populations, valley-dependent energy shifts, spin-orbit coupling-induced band splitting, or Kekul\'e distortions. Here, we show that the carrier density and magnetic field dependence of the beating nodes can distinguish these mechanisms. Starting from Onsager's quantization relation, we derive scaling relations for the critical carrier density \(N_c\) for the beating nodes as a function of critical magnetic field $B_c$. A pseudomagnetic field gives \(N_c\propto B_c^2\), whereas a density-independent valley imbalance gives \(N_c\propto B_c\). A constant Dirac-band energy splitting by Zeeman-like spin-orbit coupling also gives quadratic field scaling, but with a different node sequence: \(N_{c,j}\propto(2j+1)B_{c,j}^2\) for a pseudomagnetic field and \(N_{c,j}\propto(2j+1)^2B_{c,j}^2\) for energy splitting, where \(j\) labels the beating node indices. 
These results provide quantitative constraints on different microscopic origins of valley- and spin-dependent band splittings in graphene-based systems.
\end{abstract}

\maketitle

\section{Introduction}

Magnetic quantum oscillations, such as the Shubnikov-de Haas (SdH) and de Haas-van Alphen (dHvA) effects, are among the most powerful probes of electronic structure in metals and semimetals~\cite{Hass_30, shubnikov_30, Landau_30, lifshitz1956theory, lifshitz1958theory, dingle52, Sheonberg, glazman_fermiology23, Sunit_prb22, kamal_prr20}. They arise from the successive crossing of Landau levels through the Fermi energy as the magnetic field is varied. These oscillations are typically periodic in inverse magnetic field, with a frequency determined by the extremal Fermi-surface orbit area~\cite{lifshitz1956theory, lifshitz1958theory, Glazman_prx18, Zhao_AP22}. As a result, quantum oscillations form a cornerstone of modern Fermiology~\cite{glazman_fermiology23}. While the conventional picture involves a single oscillation frequency, several mechanisms can modify this behavior. Examples include aperiodic oscillations~\cite{Rubi_npj20, rubi_2023unconventional, Slot_science23, Anamika_afm25, gao_PNAS17, Fuch_scipost18, das24}, interaction-induced anomalous oscillations~\cite{Leeb_prl21, Leeb_prb23, Leeb_prb23_diff, Huber2023}, and amplitude beating between nearby frequencies~\cite{Hatke_prb12, Ochiai2001, Saipaopan2020, leeb2025}. Beating occurs when two oscillatory components with slightly different frequencies or phases interfere, producing a slowly varying envelope with nodes where the oscillation amplitude vanishes. Since the frequency is directly related to the Fermi-surface area, the beating pattern contains information about the origin of the frequency mismatch.

In graphene-based systems, several distinct mechanisms can generate such a mismatch in quantum oscillation frequencies. A strain-induced pseudomagnetic field makes the two valleys experience different effective magnetic fields, and hence different oscillation frequencies~\cite{Guinea_10, Zeldov_Nat23, Taniguchi_acs25, Sahani_25_PMF,li2020valley,jiang2017visualizing,liu2023recent,breitwieser2017investigating,giambastiani2022electron}. Alternatively, the two oscillatory channels can acquire different Fermi-surface areas through unequal valley populations, valley-dependent energy shifts, spin-orbit coupling-induced band splittings, or Kekul\'e distortions~\cite{Das_prb89, Takatomo_prl97, Cui_16, Vasil_prb03, Peeters_prb05, islam2014beating, Tiwari_npj22, zimmerman2026}. Although these mechanisms have different microscopic origins, they can produce qualitatively similar beating envelopes, making it difficult to identify the microscopic mechanism giving rise to the beating from the oscillation pattern alone. 

\begin{figure}[t]
    \centering
    \includegraphics[width=\linewidth]{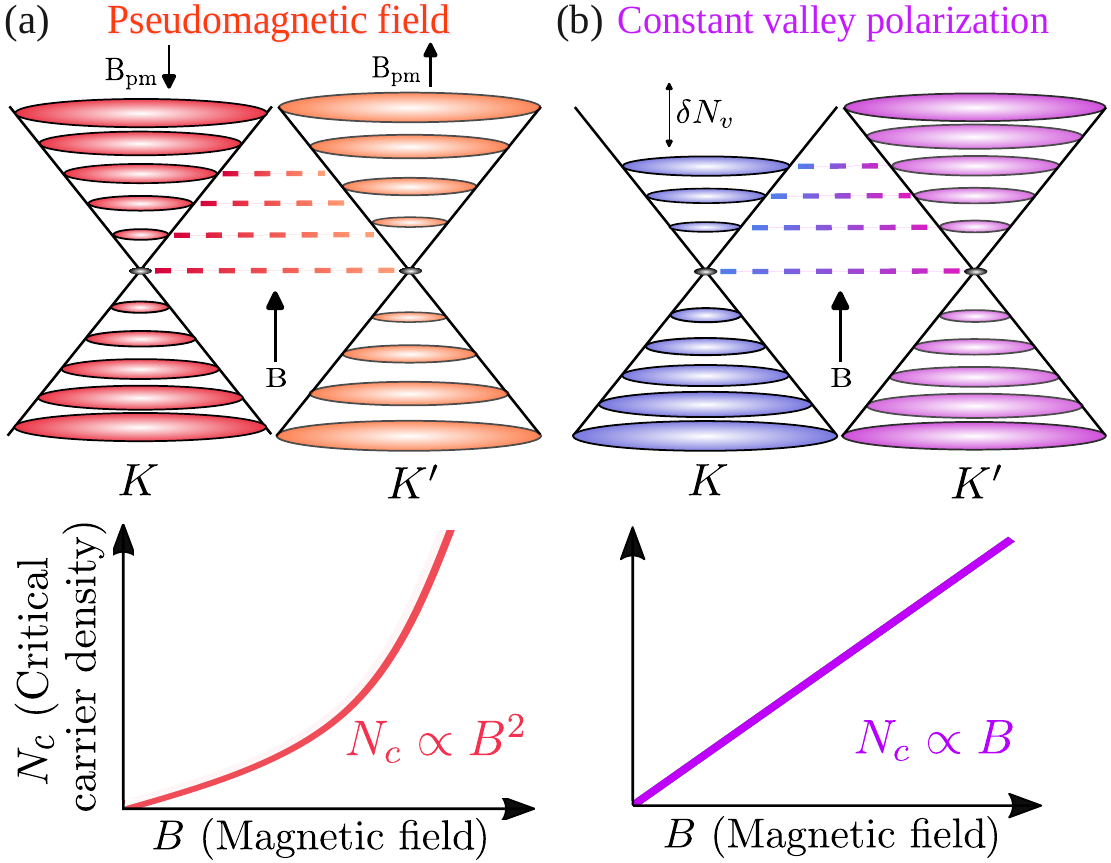}
    \caption{Schematic diagnostic of beating-node trajectories in the carrier density and magnetic field plane. The node positions are determined by the source of the frequency mismatch between the two oscillatory channels. A density imbalance proportional to carrier density, such as a constant valley polarization or a Y-type Kekul\'e velocity imbalance, gives a linear-$B$ trajectory, $N_{c,j}\propto(2j+1)B_{c,j}$. A strain-induced pseudomagnetic field gives a quadratic dependence, $N_{c,j}\propto(2j+1)B_{c,j}^2$. A constant valley- or spin-dependent splitting of the Dirac bands also gives a quadratic-$B$ dependence, but with a different node sequence, $N_{c,j}\propto(2j+1)^2B_{c,j}^2$. The combination of magnetic field and node-index dependence distinguishes the microscopic origin of the beating patterns.}
    \label{fig_scaling_schematic}
\end{figure}

This raises a natural question: how can the SdH beating spectrum be used to distinguish between different microscopic mechanisms 
that give rise to the beating pattern? A recent study showed that in strained monolayer graphene, the carrier density at the beating nodes scales quadratically with magnetic field, enabling extraction of the pseudomagnetic field from bulk transport measurements~\cite{Sahani_25_PMF}. This demonstrates that the node positions contain information beyond the visual appearance of the beating envelope and suggests a more general diagnostic framework based on the scaling of the node trajectories.

In this work, we develop such a framework using Onsager's quantization relation. We derive general scaling relations for the critical carrier density $N_{c,j}$ and magnetic field $B_{c,j}$ corresponding to the $j$th beating node, where $j=0,1,2,\ldots$ labels successive nodes. Figure~\ref{fig_scaling_schematic} shows two representative cases in the carrier density and magnetic field plane. A strain-induced pseudomagnetic field gives $N_{c,j}\propto (2j+1)B_{c,j}^{2}$, whereas a density-independent valley imbalance gives $N_{c,j}\propto (2j+1)B_{c,j}$. A constant valley- or spin-dependent energy splitting also produces a quadratic magnetic field dependence, but with a distinct node-index dependence, $N_{c,j}\propto (2j+1)^2B_{c,j}^{2}$. Thus, the magnetic field dependence provides the first diagnostic, while the node-index dependence separates mechanisms that otherwise have the same field scaling.
Table~\ref{table1} summarizes the resulting diagnostics for pseudomagnetic fields, valley imbalance, intrinsic valley splitting, valley-Zeeman spin-orbit coupling, and Y-type Kekul\'e distortion. Our results establish that, when the beating is governed by two oscillatory channels, the beating-node trajectories provide a quantitative probe of pseudomagnetic fields and valley- or spin-dependent electronic structure in graphene-based systems. 

\section{Scaling relations from Onsager's quantization condition}
\label{Sec_Onsager}

Since beating arises from the interference between two closely spaced quantum oscillation channels, Onsager's quantization condition provides a natural starting point for relating the node positions to carrier density, magnetic field, and node index. For a nondegenerate oscillatory channel with fixed spin and valley indices, the semiclassical Onsager quantization condition reads
\be
\left( n+\gamma \right)\frac{eB}{h}
=\frac{S_0(\varepsilon)}{4\pi^2}
\equiv N_0(\varepsilon)~.
\label{onsager}
\ee
Here, $n$ is the Landau level (LL) index, $S_0(\varepsilon)$ is the extremal area of the constant-energy orbit, $\gamma$ is the phase offset, and $N_0(\varepsilon)$ is the zero-field carrier density for one spin and one valley.

For a magneto-oscillatory quantity $X$, each contribution has the generic form $\delta X \propto \cos \left(2\pi n + \phi_0\right)$, where $n$ is fixed by Eq.~\eqref{onsager} and $\phi_0$ is a phase offset~\cite{Sheonberg, lifshitz1956theory, lifshitz1958theory, das24}. Beating in physical responses arises from two nearby oscillatory contributions characterized by slightly different Landau level indices, $n_1$ and $n_2$. Assuming comparable amplitudes and no additional relative phase varying between the two channels, the total response is
\bea
\delta X_{\rm tot}
&\propto&
\cos \left(2\pi n_1+\phi_0\right)
+
\cos \left(2\pi n_2+\phi_0\right)\nn \\
&=&
2\cos\left(\pi\delta n\right)
\cos\left(\pi\bar n+\phi_0\right),
\label{beat_form}
\eea
where $\delta n=n_1-n_2$ and $\bar n=n_1+n_2$. Since $\delta n \ll \bar n$, the oscillation is modulated by the slowly varying envelope $\cos(\pi\delta n)$. A beating node occurs when this envelope vanishes,
\be
|\delta n|=j+\frac{1}{2}, \qquad j=0,1,2,\cdots .
\ee
We denote the applied field and total carrier density at the $j$-th node by $B_{c,j}$ and $N_{c,j}$, respectively. The corresponding Landau level filling factor is given by $\nu_{c,j}=hN_{c,j}/(eB_{c,j})$.

The phase mismatch can arise from two physically distinct sources: a field-like asymmetry or a Fermi-surface asymmetry. To make this distinction explicit, consider graphene with two valleys, $\xi=\pm$, and allow the valleys to have different carrier densities. We denote the spin-degenerate zero-field carrier density in valley $\xi$ by $N_\xi$, so that $N_\xi=2N_0$ and the total carrier density is $N=N_+ + N_-$. We will use this notation for total density, and spin summed density for each valley, throughout the manuscript. We also allow the valleys to experience different effective magnetic fields,
\be
B_\xi=B-\xi B_{\rm pm},
\ee
where $B_{\rm pm}$ is a valley-contrasting pseudomagnetic field (PMF). Such PMFs can arise from nonuniform strain in hexagonal two-dimensional systems~\cite{Zeldov_Nat23, Taniguchi_acs25, Sahani_prl24}.

The Onsager condition for the spin-summed density in valley $\xi$ is
\be
n_\xi+\gamma
=\frac{hN_\xi}{2e(B-\xi B_{\rm pm})}
\approx
\frac{hN_\xi}{2eB}\left(1+\xi\frac{B_{\rm pm}}{B}\right),
\label{gen_onsager}
\ee
valid to leading order in $B_{\rm pm}/B$, where $n_\xi$ denotes the LL index for valley $\xi$. We have assumed that the two channels have the same phase offset $\gamma$. A constant relative offset shifts the node indexing but does not change the leading field or density powers. Subtracting the two valley quantization conditions gives
\be \label{gen_scaling}
\delta n
=\frac{h}{2eB}
\left[
\delta N
+N\frac{B_{\rm pm}}{B}
\right],
\ee
where $\delta N=N_+-N_-$ is the valley carrier density imbalance. The two terms in Eq.~\eqref{gen_scaling} have different physical origins. The term $N B_{\rm pm}/B$ is field-like: it appears when the two valleys experience different effective magnetic fields. The term $\delta N$ is Fermi-surface-like: the two channels have different carrier densities and therefore different SdH frequencies even without a field-like asymmetry.

In this illustration, the Fermi-surface asymmetry comes from differential valley population. More generally, the same density-imbalance term can represent valley-dependent band shifts, spin-orbit-induced spin splitting, velocity imbalance, or multiple bands contributing as distinct oscillatory channels. These realizations are organized in Sec.~\ref{Sec_VP_graphene}. We now focus on the beating-node scaling when only one asymmetry, either field-like or Fermi-surface-like, is present. 

\subsection{Field-like asymmetry}

Long-wavelength nonuniform strain generates a valley-contrasting pseudomagnetic field in graphene-based hexagonal systems, which preserves global time-reversal symmetry. As a result, carriers in the two valleys experience effective magnetic fields $B\pm B_{\rm pm}$. In these systems, there is no zero-field Fermi-surface asymmetry from valley or spin splitting, so $\delta N=0$. The phase mismatch responsible for beating therefore originates solely from field-like asymmetry. Applying the node condition to Eq.~\eqref{gen_scaling} gives
\be \label{quadratic_B_scaling}
N^{\rm PMF}_{c,j}
=\frac{e}{h|B_{\rm pm}|}(2j+1)B_{c,j}^2~.
\ee
Thus, a PMF-induced field-like asymmetry produces a quadratic magnetic field dependence and a linear node-index dependence in the critical density,
$N^{\rm PMF}_{c,j}\propto (2j+1)B_{c,j}^2$.
This scaling provides a direct way to extract the magnitude of the pseudomagnetic field from beating-node positions in non-uniformly strained graphene~\cite{Sahani_25_PMF}.

\subsection{Fermi-surface asymmetry}\label{FS_assy}
In the absence of PMF, beating in SdH oscillations originates only from unequal Fermi-surface areas. Then Eq.~\eqref{gen_scaling} reduces to
\be
|\delta N(N_{c,j})|
=\frac{e}{h}(2j+1)B_{c,j}~.
\label{FS_node_condition}
\ee
The field dependence of the node trajectory is therefore controlled by how the density imbalance depends on the total carrier density. This density, or Fermi-surface, asymmetry can arise from SOC or from valley splitting. The two channels experience the same applied field, but their extremal orbit areas differ. In several cases, the density imbalance itself depends on the total density. For generality, we write $|\delta N(N)|=A N^r$, 
where $A$ and $r$ are material- and mechanism-specific parameters. Combining this form with Eq.~\eqref{FS_node_condition}, we obtain
\be
N_{c,j}
=\left[\frac{e}{hA}(2j+1)B_{c,j}\right]^{1/r}.
\label{FS_general_scaling}
\ee
Contrasting this density-imbalance case with Eq.~\eqref{quadratic_B_scaling}, we see that within this power-law class the beating-node density scales with the node index and the magnetic field with the same power.

A useful case is a density imbalance proportional to the total density. This corresponds to an imbalance in the valley populations described by a polarization parameter $p_v\in[-1,1]$, such that $N_\xi=(1-\xi p_v)N/2$.
The magnitude of the valley population imbalance is then $|\delta N|=|p_v|N$, corresponding to $r=1$. If spin degeneracy is preserved, Eq.~\eqref{FS_general_scaling} gives
\be \label{linear_B_scaling}
N^{\rm VP}_{c,j}
=\frac{e}{h|p_v|}(2j+1)B_{c,j}~.
\ee
A concrete example of this constant-polarization scaling is Y-type Kekul\'e distortion in graphene~\cite{Eom_nano20,Christopher_NP16,Sahani_25_PMF}, which produces a Fermi-velocity imbalance between the two inequivalent valleys. See Sec.~\ref{Sec_FS_r1} for details. 

In contrast, a constant energy splitting of two Dirac valleys gives $|\delta N|\propto N^{1/2}$. For example, if the two Dirac valleys are shifted by energies $\pm \Delta_0$, the resulting carrier density imbalance is $|\delta N|\simeq \dfrac{2|\Delta_0|}{\sqrt{\pi}\hbar v_F}N^{1/2}$, as discussed in Sec.~\ref{Sec_FS_rhalf}. Substituting $r=1/2$ in Eq.~\eqref{FS_general_scaling} yields
\be
N_{c,j}\propto (2j+1)^2B_{c,j}^2~.
\ee
Although both the PMF-induced and valley-contrasting energy-splitting trajectories are quadratic in $B_{c,j}$, their node-index factors are different. A PMF gives $(2j+1)B_{c,j}^2$, whereas a Dirac energy splitting gives $(2j+1)^2B_{c,j}^2$. This distinction offers a direct way to identify the microscopic origin of the observed beating in experiments.

\begin{figure*}[t]
    \centering
    \includegraphics[width=\linewidth]{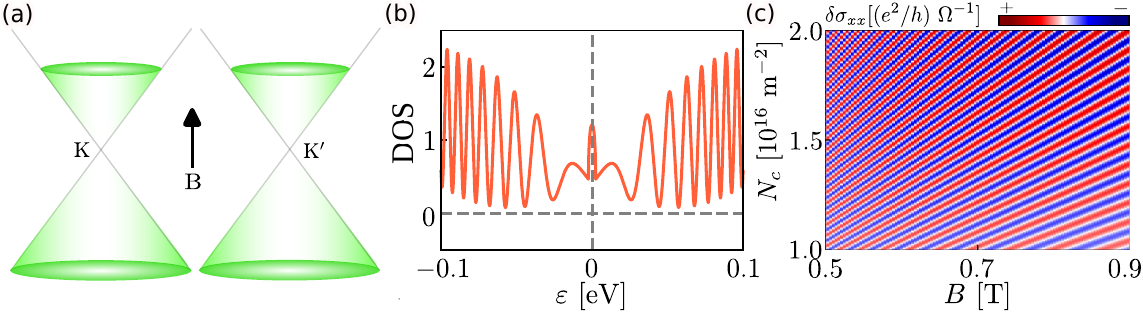}
    \caption{Single-frequency quantum oscillations in pristine single-layer graphene. (a) The two valleys have identical Fermi-surface areas and experience the same applied field. (b) The density of states shows the usual SdH oscillations as a function of energy. We use $B=1~{\rm T}$, $v_F=10^6~{\rm m/s}$, and $\Delta=0$. (c) The oscillatory longitudinal magnetoconductivity $\delta\sigma_{xx}$ in the $N$-$B$ plane contains no beating nodes because the two valley contributions have the same frequency. We use $n_{\rm im}=10^{13}~{\rm m^{-2}}$ and $k_s=10^8~{\rm m^{-1}}$.}
    \label{fig_2}
\end{figure*}

\subsection{Universal scaling ratio}
\label{Onsager_univ_rat}

The same two-channel interference picture also gives a simple mechanism-independent relation. The rapid SdH oscillations are controlled by the average phase of the two channels, whereas the beating envelope is controlled by their phase difference; see Eq.~\eqref{beat_form}. The number of fast oscillations between neighboring beating nodes is therefore set by the ratio of the average LL index to the LL-index mismatch.

At the $j$-th beating node, the envelope condition is $|\delta n|=j+\frac{1}{2}$. The local number of full rapid oscillations between neighboring beating nodes is
\begin{equation}
\mathcal{N}_{{\rm osc},j}
=
\frac{\bar n}{2|\delta n|}
=
\frac{n_1+n_2}{2j+1}.
\label{Nosc_universal}
\end{equation}
The factor of two in the denominator appears because the interval between two neighboring envelope nodes corresponds to a half-period of the beat envelope.

For the two-channel beating considered in this work, the filling factor counts the total number of occupied LLs from both channels. With spin degeneracy included, this gives
\begin{equation}
\nu_{c,j}=2(n_1+n_2).
\end{equation}
Combining this with Eq.~\eqref{Nosc_universal}, we obtain the universal ratio
\begin{equation}
\frac{\nu_{c,j}}{\mathcal{N}_{{\rm osc},j}}
=
2(2j+1).
\label{universal_ratio}
\end{equation}
This ratio is universal in the sense that it does not depend on whether the beating originates from a PMF, valley polarization, intrinsic valley splitting, SOC, or Kekul\'e distortion. The microscopic mechanism determines how $N_{c,j}$ and $B_{c,j}$ scale, but these details cancel in Eq.~\eqref{universal_ratio}. The ratio, therefore, does not identify the mechanism by itself. Instead, it provides a consistency check that the observed modulation is governed by two nearby oscillatory channels with weak splitting and reliable node indexing in oscillation data. Our microscopic calculations support the same relation, as shown in Table~\ref{table1}.

We can use the above analysis as a reliable diagnostic tool under the following conditions: two dominant oscillatory channels with comparable amplitudes, weak splitting, reliable node indexing, slowly varying damping factors and smooth prefactors, and no large density inhomogeneity. Outside this regime, the node trajectory can still indicate the source of beating, but it should be interpreted more carefully.

\section{Quantum oscillations in graphene}\label{Mag_osc}

We first develop the reference oscillatory response in monolayer graphene in the absence of any field-induced or Fermi-surface asymmetry. This will serve as a baseline for comparing with the cases with beating nodes later. For a single valley of gapped monolayer graphene, the low-energy Hamiltonian is
\be \label{Ham_dirac}
{\cal H}^\xi=\hbar v_F(\xi k_x\sigma_x+k_y\sigma_y)+\Delta\sigma_z~,
\ee
where $\xi=\pm$ labels the valleys~\cite{ghorai2024}, $v_F$ is the Fermi velocity, and $\Delta$ is a staggered sublattice potential. The finite $\Delta$ allows us to keep the valley-dependent zeroth Landau level explicit, while the main scaling results below are evaluated in the gapless limit when appropriate. In a uniform magnetic field, we substitute $\hbar{\bm k}\rightarrow\hbar{\bm k}+e{\bm A}$, with ${\bm B}=\boldsymbol{\nabla}\times{\bm A}$. The Landau levels are
\bea\label{LL_disp}
\varepsilon_n =
\begin{cases}
\lambda \sqrt{\Delta^2+2n\hbar^2\omega_c^2}~, & n\neq 0~,\\
-\xi\Delta~, & n=0~,
\end{cases}
\eea
where $\lambda=\pm1$, $\omega_c=v_F/l_B$, and $l_B=\sqrt{\hbar/eB}$. In the gapless limit used for the main scaling discussion, $\Delta=0$, the two valleys have identical spectra and identical Fermi-surface areas; see Fig.~\ref{fig_2}(a).

\begin{figure*}[t]
    \centering
    \includegraphics[width=\linewidth]{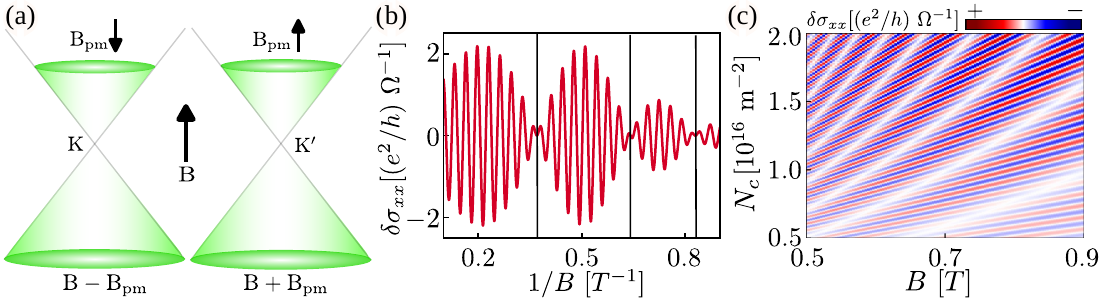}
    \caption{Field-like beating from a valley-contrasting pseudomagnetic field. (a) The two valleys have equal carrier density but experience $B_\xi=B-\xi B_{\rm pm}$. (b) The oscillatory response as a function of $1/B$ shows SdH oscillations modulated by a PMF envelope. (c) In the $N$-$B$ plane, beating nodes follow $N_{c,j}^{\rm PMF}\propto(2j+1)B_{c,j}^2$. We use $B_{\rm pm}=60~{\rm mT}$.}
    \label{fig_3}
\end{figure*}

Disorder broadens the Landau levels and produces a smooth oscillatory density of states. Keeping the first harmonic, the spin-summed DOS in valley $\xi$ can be written as
\be
D_\xi(\varepsilon)=D_{\rm LL0}+D_0(\varepsilon)
\left[1+2\Omega_D(\varepsilon,B)
\cos\left(\frac{2\pi F(\varepsilon)}{B}\right)\right]~.
\label{DOS_osc}
\ee
The detailed derivation is given in Appendix~\ref{app_A}. Here,
$D_{\rm LL0}=\dfrac{1}{\pi^2 l_B^2 \Gamma_0^2}
\sqrt{\Gamma_0^2-\dfrac{(\varepsilon+\Delta)^2}{4}}~\Theta\left(2\Gamma_0-|\varepsilon+\Delta|\right)$
denotes the zeroth-LL contribution. Here, $\Theta(x)=1$ for $x>0$ and $0$ otherwise is the step function, and $\Gamma_0$ is the impurity-induced LL broadening parameter. The zero-field DOS per spin-degenerate valley is $D_0(\varepsilon)= |\varepsilon|/( \pi \hbar^2 v_F^2)$. The factor
$\Omega_D(\varepsilon,B)=\exp [- 2 \pi^2 ( \Gamma_0\varepsilon)^2/(\hbar \omega_c)^4 ]$
accounts for damping due to LL broadening. Equation~\eqref{DOS_osc} contains a zeroth-LL contribution and the oscillatory contribution from the nonzero Landau levels. The oscillatory contributions are periodic in $1/B$, with frequency $F(\varepsilon)=(\varepsilon^2-\Delta^2)/(2e\hbar^2v_F^2)$. 
Figure~\ref{fig_2}(b) shows the corresponding Landau level quantum oscillations in the DOS as a function of energy.

The longitudinal magnetoconductivity has the same oscillatory phase. Starting from the collisional conductivity and including finite-temperature damping, as summarized in Appendix~\ref{App_B}, the valley-resolved result is
\be \label{sigma_osc_sdh}
\begin{aligned}
\sigma_{xx}^\xi(N_\xi,B_\xi)
&\approx
\sigma_0(N_\xi,B_\xi)
\\
&\quad\times
\left[
1+2\Omega_T\Omega_D
\cos\left(\frac{2\pi F(N_\xi)}{B_\xi}\right)
\right]~,
\end{aligned}
\ee
where
\be
\begin{aligned}
\sigma_0(N_\xi,B_\xi)
&=
\frac{e^2}{h}\frac{n_{\rm im}U_0^2}{2k_s^2\Gamma_0}
\frac{2N_\xi}{e\hbar v_F^2B_\xi}
\\
&\quad\times
\frac{\pi\hbar^2v_F^2N_\xi+2\Delta^2}
{\sqrt{2\pi\hbar^2v_F^2N_\xi+\Delta^2}}.
\end{aligned}
\ee
Here, $N_\xi=(\mu^2-\Delta^2)/(2\pi\hbar^2v_F^2)$ is the spin-summed carrier density in valley $\xi$, and therefore $F(N_\xi)=hN_\xi/(2e)$. The parameters $U_0 = e^2/(2 \epsilon_0 \epsilon_r)$, $n_{\rm im}$, $k_s$, $\epsilon_0$, and $\epsilon_r$ denote the screened impurity strength, impurity density, screening wave vector, vacuum permittivity, and relative permittivity, respectively. The factor $\Omega_T=(T/T_D)/\sinh(T/T_D)$ is the thermal damping factor, while $\Omega_D$ is the disorder-broadening factor.

In pristine graphene, $N_+=N_-=N/2$ and $B_+=B_-=B$. In this case, the two valley contributions have the same phase $2\pi F(N/2)/B$ and add without producing a beating pattern, 
\be
\sigma_{xx}(N,B)=\sum_{\xi=\pm}\sigma_{xx}^\xi(N/2,B)~.
\ee
Figure~\ref{fig_2}(c) shows the oscillatory part of the longitudinal magnetoconductivity,
$\delta \sigma_{xx}=\sigma_{xx}-2\sigma_0(N/2,B)$ in the $B$-$N$ plane. The alternating red and blue stripes denote successive SdH oscillations. Since the two valley channels have the same oscillation frequency in this reference case, no beating nodes appear. This provides the baseline against which the beating patterns generated by field-like or Fermi-surface asymmetries are compared below.

\section{Field-like asymmetry: pseudomagnetic field}\label{Sec_PMF_graphene}

Long-wavelength strain in graphene acts as a valley-dependent gauge field and generates a pseudomagnetic field~\cite{Guinea_10}. For a honeycomb lattice, the strain-induced vector potential is
\be
\bm{A}_{\rm pm}
=\frac{\hbar\beta\kappa}{2a_0}(u_{xx}-u_{yy},-2u_{xy})~,
\ee
where $\beta\approx2$--$3$ is the Gr\"uneisen parameter~\cite{Kitt2012,deJuan2012}, $a_0$ is the lattice constant, and $\kappa\approx1/3$ describes the elastic response. The corresponding pseudomagnetic field ${\bm B}_{\rm pm}=\boldsymbol{\nabla}\times\bm{A}_{\rm pm}$ couples with opposite sign to the two valleys. It therefore preserves global time-reversal symmetry, while making the effective magnetic field valley dependent.

For smooth strain profiles whose correlation length exceeds the cyclotron radius, the PMF can be treated as locally uniform along an orbit~\cite{Sahani_25_PMF}. For such cases, strain enters the oscillatory response through valley-dependent magnetic field $B_\xi=B-\xi B_{\rm pm}$, implying valley-dependent Landau quantization; see Fig.~\ref{fig_3}(a). Consequently, the low-energy Hamiltonian can be written in terms of the valley-dependent kinetic momentum $\bm{\Pi}_\xi=\hbar{\bm k}+e{\bm A}-e\xi{\bm A}_{\rm pm}$, where $\bm A$ is the vector potential for real magnetic field.

This is a field-like asymmetry: the two valleys have equal carrier densities, $N_\xi=N/2$, but they are quantized by different effective fields. Thus, the microscopic perturbation enters through $B_\xi$, not through a zero-field density imbalance. 
Substituting $N_\xi=N/2$ and $B_\xi=B-\xi B_{\rm pm}$ into Eq.~\eqref{sigma_osc_sdh}, and neglecting the weak $B_{\rm pm}$ dependence of the damping factors and prefactors, we obtain, 
\bea
\sigma_{xx}(N,B)
&\approx&
2\sigma_0
\bigg[
1+\Omega_T\Omega_D
\bigg\{
\cos\left(\frac{2\pi F(N/2)}{B+B_{\rm pm}}\right)\nn\\
&&+
\cos\left(\frac{2\pi F(N/2)}{B-B_{\rm pm}}\right)
\bigg\}
\bigg]~.
\label{sigma_beat_quad}
\eea
Here, $\sigma_0$, $\Omega_T$, and $\Omega_D$ are evaluated at the common density $N/2$ and applied field $B$. In the weak-PMF regime, $|B_{\rm pm}|\ll B$, this becomes
\bea
\sigma_{xx}(N,B)
&\approx&
2\sigma_0
\bigg[
1+2\Omega_T\Omega_D
\cos\left(\frac{2\pi F(N/2)}{B}\right)\nn\\
&&\times
\cos\left(\frac{2\pi F(N/2)B_{\rm pm}}{B^2}\right)
\bigg]~.
\label{sigma_beat_quad_2}
\eea
The first cosine is the rapid SdH oscillation, while the second cosine is the PMF-induced envelope, which produces the beating nodes in the SdH oscillations. Fig.~\ref{fig_3}(b) shows the resulting nodes in $\delta\sigma_{xx}$. Since its phase scales as $1/B^2$, the envelope is periodic in $1/B^2$, and the nodes are not equally spaced in $1/B$; see Fig.~\ref{fig_3}(b).

The envelope vanishes when
\be
\frac{2\pi F(N_{c,j}/2)|B_{\rm pm}|}{B_{c,j}^2}
=\left(j+\frac{1}{2}\right)\pi .
\ee
Using $F(N/2)=hN/(4e)$ gives a node trajectory controlled by the magnitude of the PMF,
\be \label{n_c_pmf}
N^{\rm PMF}_{c,j}
=\frac{e}{h|B_{\rm pm}|}(2j+1)B_{c,j}^2~.
\ee
Note that the above expression exactly matches that obtained from Onsager's quantization relation, see Eq.~\eqref{quadratic_B_scaling}. Figure~\ref{fig_3}(c) shows $\delta \sigma_{xx}$ as a function of $B$ and $N$. Alternating blue and red stripes denote SdH oscillations, while prominent white stripes mark beating nodes with different $j$ values. The nodes follow $N_c \propto B_j^2$ in the $N$-$B$ plane. Tracking this trajectory in oscillatory magnetoresistance enables bulk extraction of $B_{\rm pm}$ and has recently revealed millitesla-scale pseudomagnetic fields in strained graphene~\cite{Sahani_25_PMF}.

The associated critical filling factor and local oscillation count are
\be
\nu^{\rm PMF}_{c,j}
=\frac{hN^{\rm PMF}_{c,j}}{eB_{c,j}}
=\frac{(2j+1)B_{c,j}}{|B_{\rm pm}|},
\qquad
\mathcal{N}^{\rm PMF}_{{\rm osc},j}
=\frac{B_{c,j}}{2|B_{\rm pm}|}.
\ee
The oscillation count increases linearly with $B$, consistent with Fig.~\ref{fig_3}(b). Hence $\nu_{c,j}^{\rm PMF}/\mathcal{N}^{\rm PMF}_{{\rm osc},j}=2(2j+1)$, as required by the two-channel ratio in Sec.~\ref{Onsager_univ_rat}.
The PMF node condition carries a single power of $(2j+1)$ multiplying $B_{c,j}^2$. This node-index factor will differ from the Dirac energy-splitting case below, even though both trajectories are quadratic in $B_{c,j}$.

The same PMF scaling extends to bilayer graphene. Although monolayer graphene has $N\propto\mu^2$ and bilayer graphene has $N\propto\mu$, the beating condition is expressed in carrier density, which is always proportional to the Fermi-surface area in two dimensions. Equation~\eqref{n_c_pmf} therefore remains valid for the bilayer model treated in Appendix~\ref{App_C} to leading order in the PMF.

\section{Fermi-surface asymmetry driven beating}
\label{Sec_VP_graphene}

We now turn to beating generated by Fermi-surface asymmetry. In this case, the two oscillatory channels experience the same applied magnetic field, but their extremal Fermi-surface areas differ. Equation~\eqref{gen_scaling} then reduces to
\be
|\delta N(N_{c,j})|
=
\frac{e}{h}(2j+1)B_{c,j}~.
\label{FS_node_condition_main}
\ee
The node trajectory is therefore determined by the density dependence of the channel density imbalance $|\delta N|$. This section implements the classification developed in Sec.~\ref{FS_assy}: mechanisms with $|\delta N|\propto N$ give a linear $N_{c,j}$-$B_{c,j}$ trajectory, whereas mechanisms with $|\delta N|\propto N^{1/2}$ give a quadratic trajectory with an extra power of $(2j+1)$.

\subsection{Linear-density imbalance}
\label{Sec_FS_r1}

\subsubsection{Constant fractional imbalance}

\begin{figure*}[t]
    \centering
    \includegraphics[width=\linewidth]{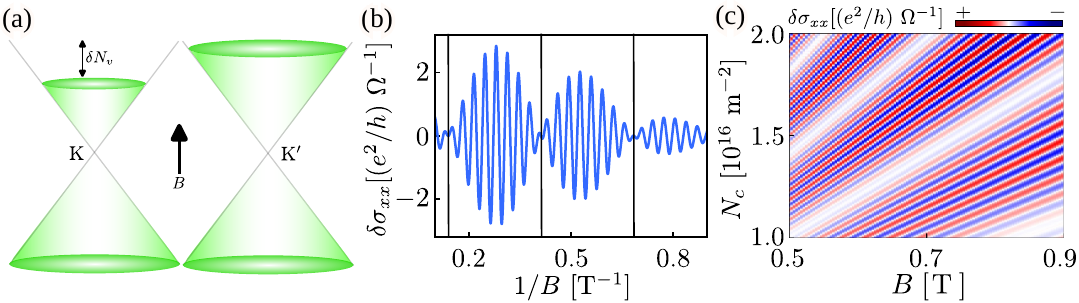}
    \caption{Fermi-surface beating from a constant fractional valley imbalance. (a) The two valleys have different carrier densities and therefore different Fermi-surface areas, while both experience the same magnetic field. (b) The oscillatory longitudinal magnetoconductivity shows a beating envelope periodic in $1/B$. (c) The node trajectory follows $N^{\rm VP}_{c,j}\propto(2j+1)B_{c,j}$. We use $p_v=0.06$.}
    \label{fig_4}
\end{figure*}

A constant fractional valley imbalance gives the simplest $r=1$ Fermi-surface asymmetry. Such an imbalance may be produced by optical pumping, such as circularly polarized excitation, that populates the valleys unequally while leaving the applied magnetic field common to both valleys. We write the spin-summed valley densities as $N_\xi=(1-\xi p_v)N/2$, where $p_v\in[-1,1]$ and $N=(N_+ + N_-)$. This gives $|\delta N|=|N_+-N_-|=|p_v|N$. In this scenario, the polarization parameter $p_v$ is independent of carrier density, as appropriate for an externally imposed nonequilibrium valley imbalance; see Fig.~\ref{fig_4}(a). Substituting $N_\xi$ into Eq.~\eqref{sigma_osc_sdh}, and evaluating the smooth prefactors and damping factors at the average density, gives
\bea \label{pop_osc}
\sigma_{xx}(N,B)
&\approx&
2\sigma_0
\bigg[
1+2\Omega_T\Omega_D
\cos\left(\frac{2\pi F(N/2)}{B}\right)\nn\\
&&\times
\cos\left(\frac{2\pi F(N|p_v|/2)}{B}\right)
\bigg]~.
\eea
The first cosine is the rapid SdH oscillation and the second is the Fermi-surface envelope. Unlike the PMF envelope in Fig.~\ref{fig_3}(b), this envelope in Fig.~\ref{fig_4}(b) is periodic in $1/B$. 
The node condition gives
\be \label{vp_linear_B_scaling}
N^{\rm VP}_{c,j}
=
\frac{e}{h|p_v|}(2j+1)B_{c,j}~.
\ee
Figure~\ref{fig_4}(c) shows the corresponding beating pattern and the linear node trajectory in the $N$-$B$ plane.
The associated filling factor and local rapid-oscillation count are
\be
\nu^{\rm VP}_{c,j}
=
\frac{2j+1}{|p_v|},
\qquad
\mathcal{N}^{\rm VP}_{{\rm osc},j}
=
\frac{1}{2|p_v|}.
\ee
Both quantities are independent of the magnetic field, while the field-independent oscillation count is also evident from Fig.~\ref{fig_4}(b). This is a useful experimental distinction from PMF-driven beating, where $\nu_{c,j}^{\rm PMF}$ and $\mathcal{N}^{\rm PMF}_{{\rm osc},j}$ grow linearly with $B_{c,j}$.


\subsubsection{Velocity imbalance}

A Y-type Kekul\'e distortion gives a second realization of the $r=1$ case. It changes the Fermi velocity in a valley-dependent way rather than shifting the two Dirac points in energy~\cite{Gamayun_njp18}. Such distortions can arise in strained graphene~\cite{Eom_nano20} or graphene grown on copper substrates~\cite{Christopher_NP16}. The spin-summed valley carrier densities are $N^{\xi}= \mu^2/(2 \pi \hbar^2 (v_F^{\xi})^2)$, where $v_F^{\xi}$ is the valley-resolved Fermi velocity. For a small velocity imbalance $\delta v_F=v_F^- -v_F^+$ with $|\delta v_F|\ll v_F$, the resulting density imbalance is $|\delta N|
\simeq N|\delta v_F|/(v_F)$.
Therefore, the Y-type Kekul\'e case can be obtained from the constant-polarization result by the replacement $|p_v|\to|\delta v_F|/v_F$, giving
\be
N^{\rm YK}_{c,j}
=
\frac{e v_F}{h|\delta v_F|}(2j+1)B_{c,j}~.
\ee
Thus, constant fractional imbalance and Y-type Kekul\'e distortion should be diagnosed by the same linear $B_{c,j}$ and $(2j+1)$ scaling, even though their microscopic origins are different.

\begingroup
\setlength{\tabcolsep}{3.5pt}
\renewcommand{\arraystretch}{1.8}

\begin{table*}[t]
\centering
\caption{Summary of beating-node diagnostics. The phase mismatch in Eq.~\eqref{gen_scaling} is either field-like, through $NB_{\rm pm}/B$, or Fermi-surface-like, through $|\delta N|=A N^r$. The listed regimes assume weak $B_{\rm pm}/B$, constant $p_v$, small $|\delta v_F|/v_F$, $|\Delta_v|\ll\mu$, and approximately conserved $s_z$ for the valley-Zeeman case.}
\footnotesize

\resizebox{\textwidth}{!}{%
\begin{tabular}{lllll}
\hline\hline
Asymmetry class & Microscopic realization & Phase-mismatch source & $N_{c,j}(B_{c,j})$ & $\nu_{c,j}/\mathcal{N}_{{\rm osc},j}$ \\
\hline
Field-like &
PMF &
$NB_{\rm pm}/B$ &
$\dfrac{e}{h|B_{\rm pm}|}(2j+1)B_{c,j}^2$ &
$2(2j+1)$ \\[0.45em]
\hline
Fermi surface ($r=1$)&
Constant fractional valley polarization &
$|\delta N|=|p_v|N$ &
$\dfrac{e}{h|p_v|}(2j+1)B_{c,j}$ &
same \\[0.45em]
Fermi surface ($r=1$) &
Y-type Kekul\'e distortion &
$|\delta N|=N|\delta v_F|/v_F$ &
$\dfrac{e v_F}{h|\delta v_F|}(2j+1)B_{c,j}$ &
same \\[0.45em]
\hline
Fermi surface ($r=1/2$) &
Intrinsic valley splitting &
$|\delta N|\propto |\Delta_v|N^{1/2}$ &
$\dfrac{(2j+1)^2}{16\pi}\dfrac{e^2v_F^2}{\Delta_v^2}B_{c,j}^2$ &
same \\[0.45em]
Fermi surface ($r=1/2$) &
Valley-Zeeman SOC &
$|\delta N|\propto |\Delta_{\rm VZ}|N^{1/2}$ &
$\dfrac{(2j+1)^2}{16\pi}\dfrac{e^2v_F^2}{\Delta_{\rm VZ}^2}B_{c,j}^2$ &
same \\[0.45em]
\hline\hline
\end{tabular}%
}

\label{table1}
\end{table*}

\endgroup
\subsection{Square-root-density imbalance}
\label{Sec_FS_rhalf}

\subsubsection{Energy splitting of Dirac Fermi surfaces}

A constant energy splitting of two Dirac cones induced by a magnetic proximity effect gives a different density power. The splitting is fixed in energy, so the induced density imbalance grows only as the Fermi momentum. A minimal valley-dependent model is
\be \label{Ham_VP_int}
\mathcal{H}^\xi
=
\hbar v_F(\xi k_x\sigma_x+k_y\sigma_y)
+
\xi\Delta_v\sigma_0~,
\ee
where $\Delta_v$ is the valley-dependent energy shift~\cite{Justin_25,das2026}. The spin-summed carrier density in valley $\xi$ is $N_\xi=(\mu-\xi\Delta_v)^2/(2\pi\hbar^2v_F^2)$. In the limit of weak valley splitting, $|\Delta_v|\ll \mu$, a linear expansion yields a carrier density imbalance 
\be
|\delta N|
=
|N_+-N_-|
\simeq
\frac{2|\Delta_v|}{\sqrt{\pi}\hbar v_F}\sqrt{N}.
\ee
Thus, intrinsic valley splitting belongs to the $r=1/2$ Fermi-surface class. Its node trajectory is
\be
N^{\rm int}_{c,j}
=
\frac{(2j+1)^2}{16\pi}
\frac{e^2v_F^2}{\Delta_v^2}
B_{c,j}^2~.
\ee
This trajectory is quadratic in $B_{c,j}$, like the PMF trajectory, but the node-index dependence is different. A field-like PMF gives $(2j+1)B_{c,j}^2$, whereas a Dirac energy splitting gives $(2j+1)^2B_{c,j}^2$. 

\subsubsection{Valley-Zeeman spin-orbit coupling}

Valley-Zeeman spin-orbit coupling in graphene gives the same $r=1/2$ class when $s_z$ remains an approximately good quantum number. This regime is relevant for graphene proximitized by transition-metal dichalcogenides~\cite{Tiwari_prl21, Tiwari_npj22}, where the dominant proximity SOC often acts as a valley-dependent Zeeman field~\cite{Tiwari_npj22,Zollner2025SOC2D}. The low-energy Hamiltonian is
\be \label{Ham_dirac_SOC}
{\cal H}^{\xi,s_z}
=
\hbar v_F(\xi k_x\sigma_x+k_y\sigma_y)
+
\xi s_z\Delta_{\rm VZ}\sigma_0~.
\ee
The energy shift depends on $\eta=\xi s_z=\pm1$. The relevant two oscillatory channels are therefore the two $\eta$ sectors, not individual spin or valley branches. The $\eta=+1$ sector contains $(\xi,s_z)=(+,+)$ and $(-,-)$, while $\eta=-1$ contains $(+,-)$ and $(-,+)$. Each sector remains twofold degenerate, but the two sectors have different Fermi-surface areas.

Including this twofold degeneracy, the density in sector $\eta$ is $N^{\rm VZ}_{\eta}
=
(\mu-\eta\Delta_{\rm VZ})^2/(2\pi\hbar^2v_F^2)$. 
This has the same structure as constant energy splitting, with $\Delta_v$ replaced by $\Delta_{\rm VZ}$. Hence, we have
\be
N^{\rm VZ}_{c,j}
=
\frac{(2j+1)^2}{16\pi}
\frac{e^2v_F^2}{\Delta_{\rm VZ}^2}
B_{c,j}^2~.
\ee
This mapping assumes that spin remains a good quantum number, i.e., spin mixing is negligible. Strong Rashba SOC mixes spin and sublattice degrees of freedom and can modify the effective splitting, while Kane-Mele SOC mainly opens a gap and is typically weaker than the valley-Zeeman term in graphene-\ch{WSe2} devices~\cite{Tiwari_prl21, Tiwari_npj22,Zollner2025SOC2D}. Depending on the relative SOC strengths, the beating pattern can acquire corrections or be suppressed~\cite{Carlos_prr23}. Beating from spin-split bands has been observed in graphene-\ch{WSe2} heterostructures~\cite{Tiwari_npj22}.

\section{Universality of scaling relations}
\label{Sec_univ_rat}

The preceding sections show that a beating-node trajectory contains two
distinct pieces of diagnostic information. The first is the magnetic field
power. A field-like PMF produces a quadratic trajectory,
$N_{c,j}\propto B_{c,j}^2$, while a Fermi-surface imbalance with
$|\delta N|\propto N$ produces a linear trajectory,
$N_{c,j}\propto B_{c,j}$. A Fermi-surface energy splitting, for which
$|\delta N|\propto N^{1/2}$, also produces a quadratic trajectory. Thus,
the field dependence alone does not always identify the microscopic
mechanism.

The second diagnostic is the node-index dependence. It separates the two
quadratic cases, 
\be
N^{\rm PMF}_{c,j}\propto (2j+1)B_{c,j}^2,
\qquad
N^{\rm split}_{c,j}\propto (2j+1)^2B_{c,j}^2~.
\ee
Here, ``split'' denotes intrinsic valley splitting or valley-Zeeman spin
splitting in the weak-spin-mixing limit. This distinction is important:
both mechanisms give $B^2$ scaling, but they differ in the node-index dependence. 

Beyond these mechanism-dependent node trajectories, the two-channel
interference picture also gives a mechanism-independent consistency check:
\be
\frac{N_{c,j}}{\mathcal{N}_{{\rm osc},j}}
=
2(2j+1)\frac{e}{h}B_{c,j},
\qquad
\frac{\nu_{c,j}}{\mathcal{N}_{{\rm osc},j}}
=
2(2j+1).
\ee
This ratio tests
whether the observed beating is governed by two nearby oscillatory
channels with weak splitting and reliable node indexing. 

For data analysis, the diagnostic can be applied in three steps. First, extract the beating-node positions in the $N$-$B$ plane. Second, fit both the magnetic field power and the node-index dependence of $N_{c,j}(B_{c,j})$. Third, use $\nu_{c,j}/\mathcal{N}_{{\rm osc},j}=2(2j+1)$ as a two-channel consistency check and not as a mechanism identifier. The mechanism is constrained by the node trajectory. Table~\ref{table1} summarizes this diagnostic hierarchy. The table is organized according to the underlying asymmetry class and its microscopic realization. The PMF belongs
to the field-like class, whereas constant fractional valley polarization,
Y-type Kekul\'e distortion, intrinsic valley splitting, and valley-Zeeman
SOC belong to Fermi-surface classes distinguished by the density power in
$|\delta N|=A N^r$.

Within the two-channel oscillation regime, this formulation is not limited to the monolayer Dirac dispersion. The
scaling relations are expressed in terms of carrier density and
Fermi-surface area to leading order, so they remain valid whenever Onsager quantization
applies, and the oscillatory response is dominated by two nearby
frequencies. This is why the PMF scaling extends to bilayer graphene, as
verified explicitly in Appendix~\ref{App_C}.

Several effects can complicate the direct use of these diagnostics. More
than one splitting mechanism may contribute simultaneously, a strong Rashba
SOC can invalidate the simple valley-Zeeman mapping, interactions can renormalize the oscillatory spectrum, and disorder or density inhomogeneity can mimic envelope suppression. In such cases, the combined
diagnostic still constrains the origin of beating, but it should not be
read as a unique mechanism assignment. Extending this analysis to
nonlinear quantum oscillations~\cite{Sunit_prb22, Vijaysankar_prl26,
Yugui_prl26} is left for future work.

\section{Conclusion} \label{Sec_concl}
We have developed a unified framework for identifying the microscopic origin of quantum oscillation beating in graphene-based systems. Starting from Onsager’s quantization relation, we showed that the trajectories of beating nodes in the carrier density vs. magnetic field plane encode direct information about the underlying source of the phase mismatch between oscillatory channels. 
A central result of this work is that the magnetic field and node-index dependence of the beating nodes provides complementary diagnostics of the underlying mechanism. In particular, strain-induced pseudomagnetic fields, constant fractional valley imbalance, and Dirac-band energy splittings generate distinct scaling trajectories, allowing mechanisms that produce visually similar beating envelopes to be distinguished quantitatively. The resulting classification is summarized in Table~\ref{table1} and is independent of the specific transport coefficient used to detect the oscillations.

Since the analysis is formulated in terms of carrier density and Onsager quantization, it applies when the oscillatory response is governed by two nearby frequencies with well-defined node indexing. The framework, therefore, provides a practical route for constraining valley- and spin-dependent electronic structure from quantum oscillation measurements. More generally, beating-node spectroscopy can serve as a simple probe of symmetry-breaking fields and band splittings in graphene and related two-dimensional quantum materials.


\section{Acknowledgments}
A. Adhikary is supported by the Institute Fellowship, IIT Kanpur. S. Das acknowledges the Ministry of Education, Government of India, for funding support through the Prime Minister's Research Fellowship. A. Agarwal acknowledges funding from the Core Research Grant by ANRF (Sanction No. CRG/2023/007003), Department of Science and Technology, India.

\appendix

\onecolumngrid
\section{Derivation of oscillatory DOS of SLG}\label{app_A}
In this Appendix, we calculate the oscillatory DOS of monolayer graphene using the Green's function formalism, with disorder-induced broadening encoded through the self-energy. The retarded self-energy in the presence of impurity scattering satisfies
\bea\label{self_energy}
\Sigma^-(\varepsilon) = \Gamma_0^2 \sum_n \frac{1}{\varepsilon - \varepsilon_n - \Sigma^-(\varepsilon)},
\eea
where \( \Gamma_0^2 \) is the impurity scattering strength, and the sum is over all energy levels.
The resulting DOS from the imaginary part of the self-energy can be evaluated as~\cite{Zhang_prb90}

\be\label{DOS_self_energy}
D(\varepsilon)=\frac{2}{2\pi l_B^2}\,\mathrm{Im}\left[
\frac{\Sigma^{-}(\varepsilon)}{\pi \Gamma_0^2}\right].
\ee
Here, the prefactor $\dfrac{2}{2\pi l_B^2} = 2\dfrac{eB}{h}$ accounts for the Landau level degeneracy (including spin) per valley. For the zeroth Landau level, taking $\varepsilon_n = -\Delta$ in Eq.~\eqref{self_energy}, the self-energy equation reduces to a quadratic form, yielding $\Sigma^{-}(\varepsilon)=\dfrac{1}{2}\left[(\varepsilon+\Delta)\pm\sqrt{(\varepsilon+\Delta)^2 - 4\Gamma_0^2}\right]$. Substitution into Eq.~\eqref{DOS_self_energy} gives
\be
D_{\rm LL0}=\dfrac{1}{\pi^2 l_B^2 \Gamma_0^2}
\sqrt{\Gamma_0^2-\frac{(\varepsilon+\Delta)^2}{4}}~\Theta(2\Gamma_0-|\varepsilon+\Delta|)~~\text{with}~\Theta(x)=\begin{cases}
  1~~\text{for}~~x>0\\
  0~~\text{otherwise}
\end{cases}.
\ee
The zeroth LL does not contribute to the oscillatory DOS. Including contributions from non-zero Landau levels, we substitute 
$\varepsilon_n = \sqrt{2n(\hbar\omega_c)^2 + \Delta^2}$ into Eq.~\eqref{self_energy}, which yields a self-consistent equation for self-energy as $\Sigma^-(\varepsilon)=\Gamma_0^2\sum_{n} f(n)$ where $f(n) =\big(\varepsilon - \varepsilon_n - \Sigma^-(\varepsilon)\big)^{-1}$.
To evaluate the sum over $n$, we use the residue theorem to write the complex integral as 
\bea
\sum_{n=0}^\infty f(n) = -\frac{1}{2\pi i} \oint_C f(z) \pi \cot(\pi z) \, dz= - {\rm Res}[f(z)\pi \cot(\pi z)]=- \left. \frac{\pi \cot(\pi z)}{\frac{d}{dz} (\varepsilon -\varepsilon_z - \Sigma^-(\varepsilon) )} \right|_{z_0},
\eea
where the poles of $f(z)$ are given by $z_0 = \frac{(\varepsilon -\Sigma^-(\varepsilon))^2 -\Delta^2}{2 (\hbar \omega_c)^2}$ yields 
\bea 
\sum_{n=0}^\infty f(n)&=& \frac{\pi (\varepsilon -\Sigma^-(\varepsilon))}{(\hbar \omega_c)^2} \cot\left(\pi \frac{[(\varepsilon -\Sigma^-(\varepsilon))^2 -\Delta^2]}{2 (\hbar \omega_c)^2} \right).
\eea
Since $\Sigma^-(\varepsilon)$ is a complex quantity, we can express it as $\Sigma^-(\varepsilon) = \Sigma_r + i \Sigma_i$. Hence, the self-consistent expression for the self-energy is given by (assuming $\varepsilon > \Sigma^-(\varepsilon)$)
\be
\Sigma^-(\varepsilon) \approx \Gamma_0^2  \frac{ \pi \varepsilon}{(\hbar \omega_c)^2} \left( \cot(\pi n_0) \right), 
\ee
with $n_0=u-iv$. The quantities $u$ and $v$ are given by $u  = \dfrac{(\varepsilon - \Sigma_r)^2 - \Sigma_i^2 - \Delta^2}{2 (\hbar \omega_c)^2} \approx \dfrac{\varepsilon ^2 - \Delta^2}{2 (\hbar \omega_c)^2}$ and $v = \dfrac{ (\varepsilon - \Sigma_r) \Sigma_i}{ (\hbar \omega_c)^2} \approx \dfrac{ \varepsilon \Sigma_i}{ (\hbar \omega_c)^2}$.
The identity $\cot(\pi(u-i v)) = \dfrac{\sin(2\pi u) + i \sinh(2\pi v)}{ \cosh(2\pi v)-\cos(2\pi u) }$~\cite{IslamGhosh2012_modulated2DEG_SOI} gives the imaginary part of the self-energy as
\bea
{\rm Im}[\Sigma^-(\varepsilon)] & \approx & \Gamma_0^2 \frac{ \pi \varepsilon}{(\hbar \omega_c)^2}  \frac{\sinh(2\pi v)}{ \cosh(2\pi v)-\cos(2\pi u)} =\Gamma_0^2 \frac{\pi \varepsilon}{(\hbar \omega_c)^2} \left[1+ 2 \sum_{p=1}^{\infty} e^{-p 2\pi v} \cos(2\pi p u) \right]. \label{img_self}
\eea 
Here, we have used the relation $\dfrac{\sinh 2\pi v}{ \cosh 2\pi v-\cos 2\pi u } = 1 + 2 \sum_{p=1}^\infty e^{-p 2\pi v} \cos (p 2\pi u)$~\cite{IslamGhosh2012_modulated2DEG_SOI}. We retain only the first harmonic, {\it i.e.}, $p=1$. In the limit $\varepsilon \Sigma_i \ll (\hbar \omega_c)^2$, $\Sigma_i$ can be obtained iteratively. The first iteration gives $\Sigma_i = \Gamma_0^2 \pi \varepsilon/(\hbar \omega_c)^2$. Substituting Eq.~\eqref{img_self} into Eq.~\eqref{DOS_self_energy} and adding the zeroth Landau level contribution, we obtain the total DOS as follows
\be\label{DOS_osc3}
D(\varepsilon)=D_{\rm LL0}+D_0(\varepsilon)
\left[1+2 \Omega_D(\varepsilon)
\cos \left( \frac{2 \pi  F(\varepsilon)}{B} \right)\right].
\ee
where $D_0(\varepsilon)=|\varepsilon|/(\pi \hbar^2 v_F^2)$ is the zero-field DOS, $\Omega_D(\varepsilon)$ denotes disorder-induced damping, and the oscillation frequency is given by $F(\varepsilon) = (\varepsilon^2  - \Delta^2)/(2e \hbar v_F^2)$.
Equation~\eqref{DOS_osc3} gives the oscillatory density of states in a high magnetic field for a particular valley, including impurity broadening. 

\section{Derivation of finite-temperature oscillatory magnetoconductivity}\label{App_B}
This Appendix derives the finite-temperature magnetoconductivity in three steps: start from the zero-temperature collisional conductivity, rewrite the Landau-level sum as an energy integral using the oscillatory DOS, and then evaluate the thermal average that generates the Lifshitz-Kosevich damping factor.

Starting from the low-energy Hamiltonian in Eq.~\eqref{Ham_dirac} for monolayer gapped graphene, the high-field zero-temperature collisional magnetoconductivity is~\cite{das2025odd}
\bea \label{sigma_n1}
\sigma_{xx}^{\xi} = \frac{e^2}{h}\frac{n_{\rm im} U_0^2}{2\pi k_s^2 l_B^2 \Gamma_0} \sum_{n} \left[ \frac{\varepsilon_n^2-\Delta^2}{\hbar^2 \omega_c^2}\left(1+3\frac{ \Delta^2 }{\varepsilon_n^2}\right)  - \xi \frac{4\Delta}{\varepsilon_n} \right] \delta(\varepsilon_n- \mu).
\eea 
Equation~\eqref{sigma_n1} is the oscillatory magnetoconductivity from the $n\neq 0$ Landau levels. The terms proportional to $n$ produce the even-in-$B$ oscillations that survive even without valley polarization. By contrast, the term linear in $\Delta$ is tied to the valley index $\xi$ and yields an odd-in-$B$ contribution. It vanishes after summing over both valleys, but can survive in valley-polarized, time-reversal-symmetry-broken graphene, consistent with the observed odd-parity magnetoresistance in proximitized samples~\cite{Sahani_prl24}. Since this odd-in-$B$ term is much smaller than the even-in-$B$ contribution~\cite{das2025odd, Sahani_prl24}, we neglect it below and focus on the part that controls the SdH oscillations.

For the $K$ valley, we write the finite-temperature oscillatory magnetoconductivity as

\be
\sigma_{xx} = \frac{\sigma_0}{\pi l_B^2 } \sum_{n} A(\varepsilon_n)\, (-\partial_{\varepsilon} f)~~\text{with}~ ~\sigma_0 = \frac{e^2}{h} \frac{n_{\rm im} U_0^2}{2 k_s^2  \Gamma_0} ~~\text{and} ~~A(\varepsilon_n) = \frac{\varepsilon_n^2-\Delta^2}{\hbar^2 \omega_c^2}\left(1+3\frac{ \Delta^2 }{\varepsilon_n^2}\right).
\ee
The next step is to replace the discrete Landau-level sum by an energy integral weighted by the DOS. Using
$\frac{2}{2\pi l_B^2} \int dn\, g(\varepsilon_n)
= \int g(\varepsilon)\, D(\varepsilon)\, d\varepsilon$ \cite{IslamGhosh2012_modulated2DEG_SOI},
we obtain
\be
\sigma_{xx}
= \sigma_0 \int A(\varepsilon)\, D_0(\varepsilon)\,
\left[1+2 \Omega_D(\varepsilon)
\cos\!\left( \frac{2 \pi  F(\varepsilon)}{B} \right)\right] (-\partial_\varepsilon f)\, d\varepsilon.
\ee
To isolate the temperature dependence, we assume that $A(\varepsilon)$, $D_0(\varepsilon)$, and $\Omega_D(\varepsilon)$ vary slowly on the scale $k_B T$ compared to the oscillatory phase. We therefore evaluate them at $\varepsilon=\mu$ and pull them out of the integral,

\be
\sigma_{xx} \approx \sigma_0\, A(\mu)\, D_0(\mu)
\left[1+2 \Omega_D(\mu) I(T)\right],
\ee
where all temperature dependence is contained in
\be
I(T)=\int_{-\infty}^{\infty} d\varepsilon\,
\left(-\partial_\varepsilon f\right)
\cos\!\left( \phi \right), \qquad \phi=\frac{2 \pi F(\varepsilon)}{B}.
\ee
Expanding the phase around the chemical potential, $\phi(\varepsilon)=\phi(\mu)+(\varepsilon-\mu)\, \phi'(\mu)$, and dropping the sine term, we obtain
\bea
I(T)&=&\cos \left[\phi(\mu)\right]
\int_{-\infty}^{\infty}
\frac{\beta_T \cos[(\varepsilon-\mu)\phi'(\mu)]}
{4\cosh^2 \left(\beta_T(\varepsilon-\mu)/2\right)}\, d\varepsilon,
\eea
where $\beta_T = 1/(k_B T)$ is the inverse temperature. Using the substitution $y=\frac{\beta_T(\varepsilon-\mu)}{2}$, with $d\varepsilon=\frac{2}{\beta_T}dy$, we have
\bea
I(T)&=& 2\cos \left[\phi(\mu)\right]
\int_{-\infty}^{\infty} \frac{\cos \left[\frac{2\phi'(\mu)}{\beta_T}y\right]}{4\cosh^2 y}\, dy = 
\frac{T/T_D}{\sinh(T/T_D)}\, \cos \left[\phi(\mu)\right].
\eea
Here, we used the standard integral $\int_{-\infty}^{\infty} \frac{\cos(ay)}{\cosh^2 y}\, dy
= \frac{\pi a}{\sinh\left(\pi a/2\right)}$. This step produces the familiar thermal damping factor, with Dingle temperature $T_D = (\hbar\omega_c)^2/(2\pi^2  k_B \mu)$. The finite-temperature conductivity for valley $\xi=+1$ then becomes
\be
\sigma_{xx} \approx \sigma_0\, A(\mu)\, D_0(\mu)
\left[1+2 \frac{T/T_D}{\sinh(T/T_D)} \Omega_D(\mu)  \cos \left[\phi(\mu)\right]\,
\right].
\ee
The magnetoconductivity for valley $\xi$ is then
\bea
\sigma_{xx}^\xi \approx  \frac{e^2}{h}\frac{n_{\rm im} U_0^2}{2k_s^2  \Gamma_0} \frac{\mu}{\pi (\hbar v_F)^2} \left[ \frac{\mu^2 -\Delta^2}{2(\hbar 
\omega_c)^2}\left(1+3\frac{\Delta^2}{\mu^2}\right)  \right]  \left[1+ 2  \Omega_T \Omega_D(\mu) \cos\left( \frac{2\pi F(\mu)}{ B}\right) \right] . \label{sigma_osc_mu}
\eea 
Here, $\Omega_T = \dfrac{T/T_D}{\sinh(T/T_D)}$ is the temperature-dependent broadening factor. This is the Lifshitz-Kosevich form quoted in the main text. Substituting $N_\xi=(\mu^2-\Delta^2)/(2 \pi \hbar^2 v_F^2)$, $F(N_\xi)=h N_\xi/(2 e)$, and allowing the field to be valley dependent, $B\to B_\xi$, gives the single-valley expression

\bea
\sigma_{xx}^\xi \approx \frac{e^2}{h}\frac{n_{\rm{im}} U_0^2}{2 k_s^2 \Gamma_0} \frac{2N_\xi}{ e \hbar v_F^2 B_\xi} \frac{\pi \hbar^2 v_F^2 N_\xi + 2\Delta^2}{\sqrt{2 \pi \hbar^2 v_F^2 N_\xi + \Delta^2}}\bigg[1+ 2 \Omega_T \Omega_D \cos\left( \frac{2\pi F(N_\xi)}{ B_\xi}\right)  \bigg] . \label{sigma_osc}
\eea 
This expression is the characteristic Lifshitz-Kosevich form of quantum oscillations in monolayer graphene and is the starting point for the beating analysis in the main text.

\section{Derivation of scaling relation in BLG}\label{App_C}

This Appendix extends the strain-induced PMF scaling analysis to BLG. We first review the Landau levels in the low-energy Hamiltonian and then derive the oscillatory DOS, magnetoconductivity, and resulting scaling behavior.

Starting from the full four-band tight-binding description of Bernal-stacked bilayer graphene~\cite{ghorai2025}, projection onto the two low-energy bands near charge neutrality gives~\cite{McCann2006_bilayerQHE, Fuch_scipost18}
\begin{equation}
    {\cal H} = -\frac{1}{2m}\left[\left(p_x^2 - p_y^2\right)\sigma_x+ \xi\,(p_y p_x + p_x p_y)\,\sigma_y\right]+ \Delta\,\sigma_z ,
\end{equation}
where $m=\gamma_1/(2 v_F^2)$ is the effective mass, $\gamma_1$ denotes the interlayer dimer hopping parameter, $v_F$ is the Fermi velocity of monolayer graphene, and $\Delta$ denotes a sublattice potential. In the presence of a magnetic field ($\bm{B}=B\hat{z}$), the momentum $\bm p$ is replaced by $(\bm{p}+e\bm{A})$. Under the Landau gauge $\bm{A} = (0,xB,0)$, the energy eigenvalues are given by
\begin{equation}
\varepsilon_n =
\begin{cases} 
\lambda \sqrt{\Delta^2 + \hbar^2 \omega_c^2 n(n-1)}, & n \neq 0,1, \\ 
-\xi\Delta, & n = 0,1.
\end{cases}
\end{equation}
where $\omega_c=eB/m$ is the cyclotron frequency. Substituting $\cos{\alpha}=\Delta/\varepsilon_n$, the wavefunction for $n \neq 0,1$ can be written as
\begin{equation}
   \Phi_{n,k_y}^{\xi=1}(X) = \frac{e^{ik_y y}}{\sqrt{L_y}}
\begin{pmatrix}
    \cos{(\alpha/2)} \phi_{n-2}(X) \\
    \sin{(\alpha/2)} \phi_n(X)
\end{pmatrix}, 
\Phi_{n,k_y}^{\xi=-1}(X) = \frac{e^{ik_y y}}{\sqrt{L_y}}
\begin{pmatrix}
    \cos{(\alpha/2)} \phi_{n}(X) \\
    \sin{(\alpha/2)} \phi_{n-2}(X)
\end{pmatrix}. \label{wavefun} 
\end{equation}
For the beating pattern, we work in the high-Landau-level regime, where many distinguishable LLs are occupied. The energy spectrum then simplifies to $\varepsilon_n \approx \sqrt{\Delta^2 + n^2 \hbar^2 \omega_c^2 }$, as reported in Ref.~[\onlinecite{Zeldov_Nat23}]. 

%
Following the monolayer analysis, we evaluate the oscillatory DOS in BLG using the self-energy formalism. The corresponding self-energy follows from Eq.~\eqref{self_energy}, where poles of $f(z)$ are given by $z_0 = \sqrt{(\varepsilon -\Sigma^-(\varepsilon))^2 -\Delta^2}/(\hbar \omega_c)$. 
For simplicity, we take $\Delta=0$, giving
\be\label{self_energy_2}
\Sigma^-(\varepsilon) \approx \Gamma_0^2  \frac{ \pi }{\hbar \omega_c} \left( \cot(\pi n_0) \right), 
\ee
where  $n_0=u-iv$ with $u = (\varepsilon - \Sigma_r)/ \hbar \omega_c$ and $v = \Sigma_i/\hbar \omega_c$. 
Substituting Eq.~\eqref{self_energy_2} into Eq.~\eqref{DOS_self_energy}, we obtain the density of states
\bea
D(\varepsilon) &=& D_0 \left[1+ 2  \Omega_D\cos\left( \frac{2\pi F(\varepsilon)}{B}\right) \right] . \label{DOS_osc2}
\eea 
where $D_0 = m/(\pi \hbar^2)$ denotes the zero-field DOS, $\Omega_D=\exp\left(-\dfrac{2\pi^2\Gamma_0^2}{\hbar^2\omega_c^2}\right)$ is the damping factor arising from Landau-level broadening, and $F(\varepsilon)=m\varepsilon/(e\hbar)$ represents the oscillation frequency. Equation~\eqref{DOS_osc2} gives the oscillatory density of states in a high magnetic field for a particular valley.

To obtain the finite-temperature magnetoconductivity and beating behavior, we start from the zero-temperature oscillatory magnetoconductivity for BLG. The longitudinal conductivity arises mainly from charged-impurity scattering of cyclotron orbits, giving the collisional conductivity
\begin{equation}\label{final}
    \sigma_{xx}=C(B)
\sum_{n} \left[(4n-2)\left(1+\frac{\Delta^2}{\varepsilon_n^2}\right) - \xi\frac{8\Delta}{\varepsilon_n} \right]\delta(\varepsilon_n - \mu).
\end{equation}
As in monolayer graphene, we can extend the analysis to the finite-temperature limit. For simplicity, we take $\Delta=0$. In the high-Landau-level regime, substituting the carrier density per valley, $N_\xi=m\mu/(\pi \hbar^2)$, the finite-temperature magnetoconductivity for valley $\xi$ becomes
\bea\label{sigma_blg}
\sigma_{xx}^\xi \approx && \sigma_0\left[1+ 2 \Omega_T \Omega_D \cos\left( \frac{2\pi F(N_\xi)}{B_\xi}\right) \right]~~\text{with} ~~\sigma_0=\frac{e^2}{h}\frac{n_{\rm im} U_0^2}{2k_s^2  \Gamma_0} \frac{4 m N_\xi}{\hbar e B_\xi}. 
\eea 
where $\sigma_0$ is the non-oscillatory contribution and $F(N_\xi)=hN_\xi/2e$ is the oscillation frequency. Here, $\Omega_T = \dfrac{T/T_D}{\sinh(T/T_D)}$ is the temperature-dependent broadening factor, with Dingle temperature $T_D = \hbar\omega_c/(2\pi^2  k_B) $. This is the Lifshitz-Kosevich form for oscillatory magnetoconductivity in graphene. 

In the presence of a valley-contrasting pseudomagnetic field, we substitute $B_\xi =B-\xi B_{\rm{pm}}$ and $N_\xi=N/2$ ($N$ is the total carrier density) into Eq.~\eqref{sigma_blg} to obtain the conductivity in each valley. Neglecting the weak $B_{\rm pm}$ dependence of the prefactors and damping factors, summing over the two valleys, $\sigma_{xx}(N, B) = \sum_\xi \sigma^{\xi}_{xx} (N/2, B_\xi)$, gives the total magnetoconductivity
\bea 
\sigma_{xx}(N,B)  \approx  2\sigma_0  \bigg[1+ 2 \Omega_T   \Omega_D   \cos\left( \frac{2\pi F(N/2)}{ B}\right)\cos\left( \frac{2\pi F(N B_{\rm{pm}} /2B ) }{B} \right) \bigg]~.
\eea
The $j$-th beating node occurs when $\cos\left( \dfrac{2\pi F(N B_{\rm{pm}} /2B ) }{B} \right) =0$, giving $h N |B_{\rm pm}|/(2 e B_{c,j}^2) = (2j + 1)/2$. In terms of total carrier density, we have 
\be 
N^{\rm{PMF}}_{c,j} = \frac{e B_{c,j}^2}{ h |B_{\rm pm}|} (2 j+1)~,
\ee
This is the same scaling obtained from Onsager quantization and from the microscopic single-layer calculation.


\twocolumngrid
\bibliography{refs}
\end{document}